\documentclass[conference]{IEEEtran}
\IEEEoverridecommandlockouts
\usepackage{cite}
\usepackage{amsmath,amssymb,amsfonts}
\usepackage{algorithmic}
\usepackage{stfloats}
\usepackage{graphicx}
\usepackage{fancyhdr}
\usepackage{booktabs}
\usepackage{textcomp}
\usepackage{xcolor}
\usepackage[T1]{fontenc}
\usepackage[utf8]{inputenc}
\usepackage{authblk}
\usepackage{fancyhdr}

\def\BibTeX{{\rm B\kern-.05em{\sc i\kern-.025em b}\kern-.08em
    T\kern-.1667em\lower.7ex\hbox{E}\kern-.125emX}}

\begin{document}
\title{Computational Performance of a Germline Variant Calling Pipeline for Next Generation Sequencing
}

\author[1]{Jie Liu}
\author[1]{Xiaotian Wu}
\author[2]{Kai Zhang}
\author[2]{Bing Liu}
\author[2]{Renyi Bao}
\author[2]{Xiao Chen}
\author[2]{\\Yiran Cai}
\author[2]{Yiming Shen}
\author[2]{Xinjun He}
\author[2]{Jun Yan}
\author[1]{Weixing Ji\thanks{Corresponding authors: Weixing Ji(jwx@bit.edu.cn) and Xinjun He(xinjun.he@yigencloud.cn).}}
\affil[1]{School of Computer Science \& Technology, Beijing Institute of Technology, Beijing, China}
\affil[2]{Nanjing YiGenCloud Institute, Nanjing, China}

\maketitle
\thispagestyle{fancy}
\cfoot{\thepage}%页脚中间
\pagestyle{fancy}

\begin{abstract}
With the booming of next generation sequencing technology and its implementation in clinical practice and life science research, the need for faster and more efficient data analysis methods becomes pressing in the field of sequencing. Here we report on the evaluation of an optimized germline mutation calling pipeline, HummingBird, by assessing its performance against the widely accepted BWA-GATK pipeline. We found that the HummingBird pipeline can significantly reduce the running time of the primary data analysis for whole genome sequencing and whole exome sequencing while without significantly sacrificing the variant calling accuracy. Thus, we conclude that expansion of such software usage will help to improve the primary data analysis efficiency for next generation sequencing.
\end{abstract}


\section{Introduction}
Next generation sequencing has revolutionarily changed the way of basic life science research and clinical practice for cancer, infectious and genetic diseases \cite{b1,b2}. Population based large scale sequencing at national level helped to collect baseline variation information for various genetic traits from healthy and susceptible individuals \cite{b3,b4,b5,b6}. With rapid increase of the output from new sequencing technologies and the gradual decrease of per base sequencing cost, whole genome sequencing and whole exome sequencing have migrated from national and multi-national research projects to daily practical clinical applications especially in the field of cancer and genetic disease diagnostics \cite{b5,b7,b8}. As it is the inherited character for next generation sequencing, the initial deposition of large amount sequencing data provided not only immediate assistances needed for its requisition, but also a rich data resource for further exploration.

BWA is a widely accepted tool for aligning sequencing reads to a reference genome \cite{b9,b10}. In particularly, BWA consists of three different algorithms: BWA-backtrack, BWA-SW and BWA-MEM. BWA-MEM is the optimized algorithm for short reads mapping. Even though BWA implements thread-level parallelism, it takes a long time for large-scale datasets. To solve this problem, many researchers used different strategies such as Hadoop, Spark, MPI to accelerate BWA \cite{b11,b12,b13}, it is still hard to reach the goal that one uses acceptable hardware resources.

The Genome Analysis Toolkit (GATK) is a set of bioinformatic tools for analyzing high-throughput sequencing (HTS) and variant call format (VCF) data \cite{b14,b15}. The toolkit is well established for germline short variant discovery from whole genome and exome sequencing data. In this study, we used GATK 4.1.2.0 as the benchmark to evaluate the performance of HummingBird pipeline.

elPrep is a high-performance tool for preparing sequencing reads for variant calling in a sequencing pipeline\cite{b16,b17,b18}. It contains optimized sequencing reads preparation steps including reads sorting, duplicates marking and base quality recalibration. elPrep prepares sequencing reads and produces identical results as SAMtools\cite{b19,b20,b21} and Picard \cite{b22}. elPrep can be run in multithread and entirely in memory, thus avoiding repeated file I/O and significantly reducing running time for sequencing reads preparation.

The power of next generation sequencing is to collect as much genetic information as possible from sequenced samples; the large amount of data generated thus require computationally intensive method by professional bioinformaticians to extract the genetic information initially looked for. However, it usually takes up to a few days to process a whole genome sequencing data and costs unacceptably long time when there are a large number of samples with standard BWA \cite{b9,b10} for alignment and the Genome Analysis Toolkit (GATK) from Broad Institute for short variant calling for primary data analysis \cite{b14,b15,b23}. Thus, more reliable, efficient and reproducible sequencing data analysis tools are needed to meet the needs of a speedy variant calling \cite{b24,b25}.

In this paper, we developed HummingBird pipeline which can reduce the clock time for germline variant calling without significantly sacrificing the variant calling accuracy. Here we evaluated a germline mutation calling pipeline, HummingBird pipeline, that was developed by YiduCloud (Beijing) company, and reported that HummingBird is a fast and accurate mutation calling pipeline that can be expanded for sequencing data analysis.

The rest of the paper is organized in this way: Section 2 compares BWA-GATK and HummingBird in methodology; Section 3 analyzes the results of BWA-GATK and HummingBird under different sample sets; the conclusions are given in Section 4.

\section{Evaluation Methodology}
HummingBird pipeline is such an endeavor developed to reduce the clock time for germline variant calling by optimizing the standard BWA-GATK software interaction with maximizing the usage of hardware. As in commonly accepted BWA-GATK pipeline, the HummingBird pipeline is built with a modified BWA alignment algorithm (HB-BWA) but preserving the underlying mathematical model. The mapped sequence reads in bam files are further piped into elPrep software \cite{b16,b17,b18}, a tool kit written in GO, that outputs sorted, duplication marked and recalibrated sequence reads, which is next processed with an optimized, C++ based GATK HaplotypeCaller (HB-HaplotypeCaller) for variant calling.

\begin{table*}[htbp]      
\centering
\caption{Information of datasets used in this study}
\label{tab1}
    \begin{tabular}{ccccc}    
        \toprule  
        Sample Name & Type & Clean Reads Files & Clean Reads \# & Clean Reads File Size\\        
        \midrule            
        NIST7035\_L001 & WES & NIST7035\_TAAGGCGA\_L001\_R1\_001\_trimmed.clean.fq.gz & 19612287 & 1.5G\\
         &  & NIST7035\_TAAGGCGA\_L001\_R2\_001\_trimmed.clean.fq.gz & 19612287 & 1.5G\\
        NIST7035\_L002 & WES & NIST7035\_TAAGGCGA\_L002\_R1\_001\_trimmed.clean.fq.gz & 19985481 & 1.5G\\
         &  & NIST7035\_TAAGGCGA\_L002\_R2\_001\_trimmed.clean.fq.gz & 19985481 & 1.6G\\
        NIST7086\_L001 & WES & NIST7086\_CGTACTAG\_L001\_R1\_001\_trimmed.clean.fq.gz & 20900829 & 1.6G\\
         &  & NIST7086\_CGTACTAG\_L001\_R2\_001\_trimmed.clean.fq.gz & 20900829 & 1.6G\\
        NIST7086\_L002 & WES & NIST7086\_CGTACTAG\_L002\_R1\_001\_trimmed.clean.fq.gz & 21242748 & 1.6G\\
         &  & NIST7086\_CGTACTAG\_L002\_R2\_001\_trimmed.clean.fq.gz & 21242748 & 1.7G\\
        SRR098401 & WES & SRR098401\_1.clean.fq.gz & 80253885 & 5.1G\\
         &  & SRR098401\_2.clean.fq.gz & 80253885 & 5.3G\\
        SRR742200 & WES & SRR742200\_1.clean.fq.gz & 24576684 & 2.0G\\
         &  & SRR742200\_2.clean.fq.gz & 24576684 & 2.0G\\
        SRR098359 & WES & SRR098359\_1.clean.fq.gz & 60017306 & 3.7G\\
         &  & SRR098359\_2.clean.fq.gz & 60017306 & 4.1G\\
        SRR768308 & WGS & SRR768308\_1.clean.fq.gz & 69675029 & 5.5G\\
         &  & SRR768308\_2.clean.fq.gz & 69675029 & 5.9G\\
        ERR091571 & WGS & ERR091571\_1.clean.fq.gz & 196150792 & 15.0G\\
         &  & ERR091571\_1.clean.fq.gz & 196150792 & 16.0G\\
        NA12878\_1000G & WGS & NA12878\_merge1\_1000G.clean.R1.fq.gz & 741582047 & 32.0G\\
         &  & NA12878\_merge1\_1000G.clean.R2.fq.gz & 741582047 & 33.0G\\
        NA12878\_NIST & WGS & NA12878\_merge2.R1.clean.fq.gz & 236287755 & 23.0G\\
         &  & NA12878\_merge2.R2.clean.fq.gz & 236287755 & 25.0G\\
        ERR262997 & WGS & ERR262997.R1.clean.fq.gz & 593216693 & 49.0G\\
         &  & ERR262997.R2.clean.fq.gz & 593216693 & 50.0G\\
        \bottomrule         
        \end{tabular}
\end{table*}

\begin{table*}[htbp]      
\newcommand{\tabincell}[2]{\begin{tabular}{@{}#1@{}}#2\end{tabular}}
\centering
\caption{Download address of datasets used in this study}
\label{tab3}
    \begin{tabular}{cc}    
        \toprule  
        Sample Name & Raw Reads Download Address\\        
        \midrule            
        NIST7035\_L001 & \tabincell{c}{ftp://ftp-trace.ncbi.nlm.nih.gov/giab/ftp/data/NA12878/Garvan\_NA12878\_HG001\_HiSeq\_Exome\\/NIST7035\_TAAGGCGA\_L001\_R1\_001\_trimmed.fastq.gz}\\
         & \tabincell{c}{ftp://ftp-trace.ncbi.nlm.nih.gov/giab/ftp/data/NA12878/Garvan\_NA12878\_HG001\_HiSeq\_Exome\\/NIST7035\_TAAGGCGA\_L001\_R2\_001\_trimmed.fastq.gz}\\
        NIST7035\_L002 & \tabincell{c}{ftp://ftp-trace.ncbi.nlm.nih.gov/giab/ftp/data/NA12878/Garvan\_NA12878\_HG001\_HiSeq\_Exome\\/NIST7035\_TAAGGCGA\_L002\_R1\_001\_trimmed.fastq.gz}\\
         & \tabincell{c}{ftp://ftp-trace.ncbi.nlm.nih.gov/giab/ftp/data/NA12878/Garvan\_NA12878\_HG001\_HiSeq\_Exome\\/NIST7035\_TAAGGCGA\_L002\_R2\_001\_trimmed.fastq.gz}\\
        NIST7086\_L001 & \tabincell{c}{ftp://ftp-trace.ncbi.nlm.nih.gov/giab/ftp/data/NA12878/Garvan\_NA12878\_HG001\_HiSeq\_Exome\\/NIST7086\_CGTACTAG\_L001\_R1\_001\_trimmed.fastq.gz}\\
         & \tabincell{c}{ftp://ftp-trace.ncbi.nlm.nih.gov/giab/ftp/data/NA12878/Garvan\_NA12878\_HG001\_HiSeq\_Exome\\/NIST7086\_CGTACTAG\_L001\_R2\_001\_trimmed.fastq.gz}\\
        NIST7086\_L002 & \tabincell{c}{ftp://ftp-trace.ncbi.nlm.nih.gov/giab/ftp/data/NA12878/Garvan\_NA12878\_HG001\_HiSeq\_Exome\\/NIST7086\_CGTACTAG\_L002\_R1\_001\_trimmed.fastq.gz}\\
         & \tabincell{c}{ftp://ftp-trace.ncbi.nlm.nih.gov/giab/ftp/data/NA12878/Garvan\_NA12878\_HG001\_HiSeq\_Exome\\/NIST7086\_CGTACTAG\_L002\_R2\_001\_trimmed.fastq.gz}\\
        SRR098401 & ftp://ftp.sra.ebi.ac.uk/vol1/fastq/SRR098/SRR098401/SRR098401\_1.fastq.gz\\
         & ftp://ftp.sra.ebi.ac.uk/vol1/fastq/SRR098/SRR098401/SRR098401\_2.fastq.gz\\
        SRR742200 & ftp://ftp.sra.ebi.ac.uk/vol1/fastq/SRR742/SRR742200/SRR742200\_1.fastq.gz\\
         & ftp://ftp.sra.ebi.ac.uk/vol1/fastq/SRR742/SRR742200/SRR742200\_2.fastq.gz\\
        SRR098359 & ftp://ftp.sra.ebi.ac.uk/vol1/fastq/SRR098/SRR098359/SRR098359\_1.fastq.gz\\
         & ftp://ftp.sra.ebi.ac.uk/vol1/fastq/SRR098/SRR098359/SRR098359\_2.fastq.gz\\
        SRR768308 & ftp://ftp.sra.ebi.ac.uk/vol1/fastq/SRR768/SRR768308/SRR768308\_1.fastq.gz\\
         & ftp://ftp.sra.ebi.ac.uk/vol1/fastq/SRR768/SRR768308/SRR768308\_2.fastq.gz\\
        ERR091571 & ftp://ftp.sra.ebi.ac.uk/vol1/fastq/ERR091/ERR091571/ERR091571\_1.fastq.gz\\
         & ftp://ftp.sra.ebi.ac.uk/vol1/fastq/ERR091/ERR091571/ERR091571\_2.fastq.gz\\
        NA12878\_1000G & ftp://ftp-trace.ncbi.nlm.nih.gov/1000genomes/ftp/phase3/data/NA12878/sequence\_read/\\
        NA12878\_NIST & \tabincell{c}{ftp://ftp-trace.ncbi.nlm.nih.gov/ReferenceSamples/giab/data/NA12878/NIST\_NA12878\_HG001\_HiSeq\_300x\\/131219\_D00360\_005\_BH814YADXX/Project\_RM8398/}\\
        ERR262997 & ftp://ftp.sra.ebi.ac.uk/vol1/fastq/ERR262/ERR262997/ERR262997\_1.fastq.gz\\
         & ftp://ftp.sra.ebi.ac.uk/vol1/fastq/ERR262/ERR262997/ERR262997\_2.fastq.gz\\
        \bottomrule         
        \end{tabular}
\end{table*}

We assessed the efficiency of HummingBird pipeline by using a serial publicly available datasets. The datasets used in this paper included 12 whole genome/exome sequencing data files from NCBI/EBI database (Table 1).  Table 2 shows the download links for the datasets. The raw reads of these datasets were processed with fastp \cite{b26} software to remove low quality reads. Figure 1 presents the major steps of HummingBird and BWA-GATK pipelines. The benchmarking BWA-GATK procedure first maps the sequencing reads by BWA MEM 0.7.17 to human hg19 reference genome, and then with SAMtools 1.9 for indexing, GATK 4.1.2.0 SortSam, MarkDuplicates, BaseRecalibrator and ApplyBQSR to perform sequencing reads sorting, duplicates marking and base recalibration. After the sequence reads have been prepared, GATK 4.1.2.0 HaplotypeCaller was used for variant calling. Both pipelines were run on a x86\_64 standalone server with Intel(R) Xeon(R) Gold 6254 CPU*2 @ 3.10GHz, 72 threads, 256GB memory and a 1.5TB SSD hard drive.

\begin{figure}[htbp]
\centerline{\includegraphics[height=5cm,width=9cm]{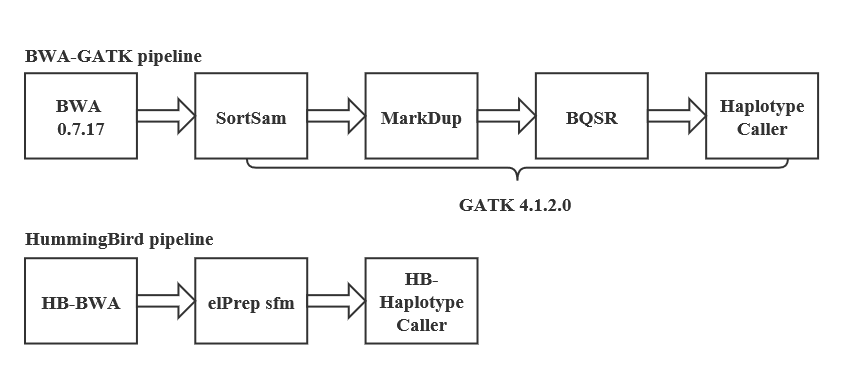}}
\caption{Major steps of BWA-GATK pipeline and HummingBird pipeline. BWA-GATK pipeline was initiated by inputting two paired-end clean reads fastq files into BWA 0.7.17. Mapped reads in sam files were next sequentially processed by GATK SortSam, MarkDuplicates, BaseRecalibrator and applyBQSR modules for sequencing reads prep. SAMtools index function was used to index variant intermediate files.  The HaplotypeCaller module of GATK 4.1.2.0 was used in last step for variant calling. HummingBird pipeline uses the same clean reads files for its modified HB-BWA, which is pipes with elPrep to sorting, duplicates marking and base recalibration. Prepared sequencing reads in bam files were further processed by the modified HB-HaplotypeCaller for variant calling.}
\label{fig1}
\end{figure}

\section{Evaluation Results}

\subsection{Running time comparison}\label{AA}
To evaluate the performance of HummingBird pipeline, we first run the pipeline along with the standard BWA-GATK procedure on the same standalone server. We evaluated the overall running time for 12 datasets downloaded from public databases. Figure 2 compares the total time cost of HummingBird pipeline and standard BWA-GATK pipeline. We found that, overall, the HummingBird pipeline processed germline mutation calling much more efficient than the standard BWA-GATK pipeline, with a maximal speedup of 12.7x for the whole genome sequencing dataset SRR768308 and 4.6x for the whole exome sequencing dataset SRR098359.

\begin{figure}[htbp]
\centerline{\includegraphics[height=6cm,width=9cm]{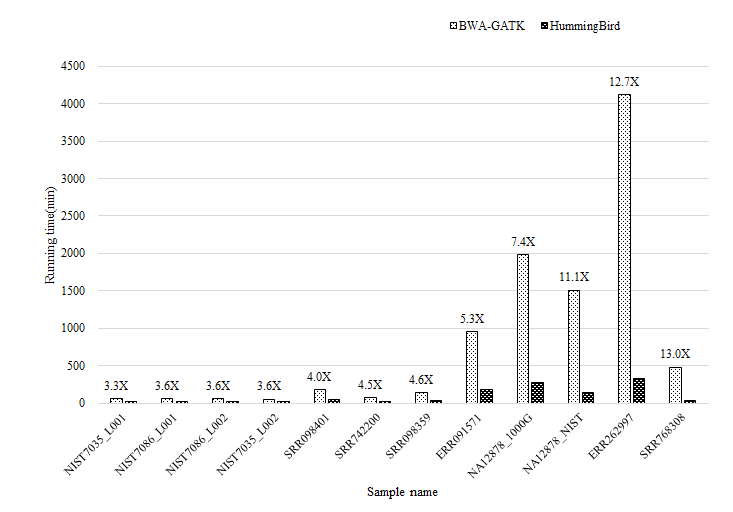}}
\caption{Total time cost of HummingBird pipeline compared with Standard BWA-GATK pipeline. Standard BWA-GATK and HummingBird pipeline were run along each other for each dataset indicated. Datasets NIST7035\_L001, NIST7035\_L002, NIST7086\_L001, NIST7086\_L002, SRR098401, SRR742200, SRR098359 are from WES sequencing, and datasets SRR768308, ERR091571, NA12878\_1000G, NA12878\_NIST, ERR262997 are from WGS sequencing.}
\label{fig2}
\end{figure}

To further assess the efficiency of individual steps along the pipe, we looked into the alignment and variant calling steps respectively as elPrep has been shown to efficiently prep mapped reads for GATK HaplotypeCaller. The time cost is shown in Figure 3A and 3B respectively. We found that HB-BWA and HB-HaplotypeCaller performed relatively more efficient than the standard BWA MEM and GATK HaplotypeCaller, with the maximal speedup of 9.4x in BWA alignment step for dataset ERR262997 and 36.9x in haplotype caller step for dataset ERR091571. 

\begin{figure}[htbp]
\centerline{\includegraphics[height=9cm,width=9cm]{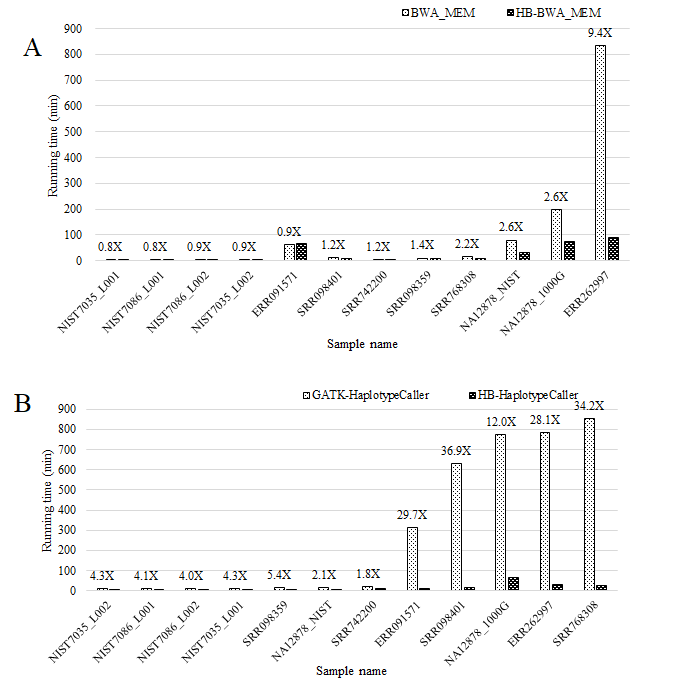}}
\caption{Time cost of alignment and haplotype caller steps of two pipelines. Running time for alignment (A) and haplotype caller (B) were recorded respectively within the complete pipelines and plotted based on duration for each dataset. }
\label{fig3}
\end{figure}

Overall, we concluded that, with the help from elPrep, the HB-BWA and HB-HaplotypeCaller within the HummingBird pipeline are more efficient for processing germline variant calling than the standard BWA-GATK procedure.

\subsection{Variant calling accuracy}
Recall reflects the percentage of variants obtained by the standard BWA-GATK benchmark pipeline that were called by HummingBird pipeline in each dataset. Precision is the percentage of called variants which match variants obtained by the standard BWA-GATK benchmark pipeline. F1-score is the harmonic mean of recall and precision, as a combined metric for evaluating overall accuracy. With the effort to assessment the accuracy of HummingBird pipeline, we calculated the result of precision and recall and further obtained F1-score on each test by the following formula:
\begin{equation}
F1 core=\frac{precision\times recall}{precision+recall}\times 2\label{eq1}
\end{equation}

All these results are shown in Figure 4. It can be seen that the precision, recall and F1-score on these data samples are consistently high across different datasets, except for dataset ERR262997, a dataset generated for large insert library evaluation, with an F1-score of 0.87.

\begin{figure*}[bp]
\centerline{\includegraphics[height=6cm,width=15cm]{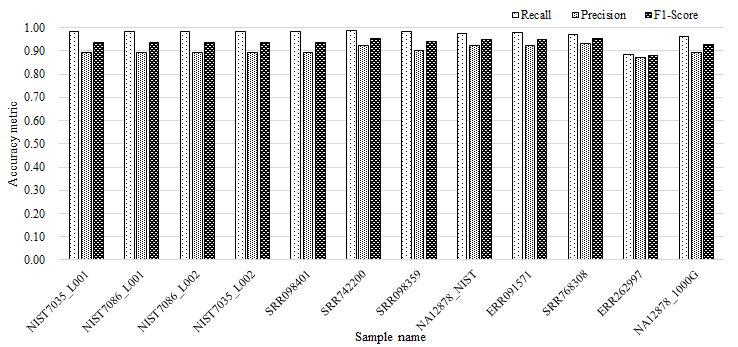}}
\caption{Accuracy of HummingBird pipeline on each sequencing. Recall, Precision and F1-Score were calculated as described for each dataset indicated. }
\label{fig4}
\end{figure*}

We also looked into the performance of HummingBird pipeline on the gold standard sample NA12878, one of the most studied samples with a better knowledge of its variants by using the hap.py evaluation tool at the high confidence intervals on chromosome 1-22, X and Y developed by Illumina Inc (19). Results on dataset NA12878-NIST showed that HummingBird pipeline generated the comparable accuracy score with standard BWA-GATK pipeline for both INDEL and SNP (Figure 5, additional results are shown in Table 3).

\begin{figure}[htbp]
\centerline{\includegraphics[height=5cm,width=9cm]{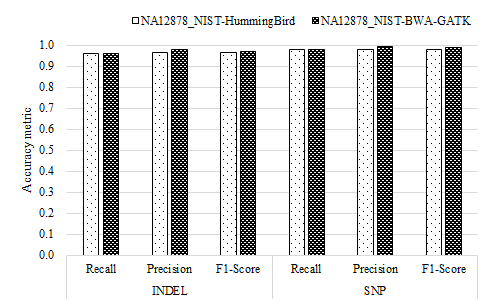}}
\caption{Accuracy comparison between HummingBird and BWA-GATK on the sample NA12878. Using the well-studied NA12878-NIST as the standard dataset, HummingBird and BWA-GATK pipeline were assessed against gold standard Illumina trustset variants (v2017-1.0)(19). INDEL and SNP were further evaluated separately for Recall, Precision and F1-Score. }
\label{fig5}
\end{figure}

\begin{table}[htbp]      
\centering
\caption{Accuracy metrics of HummingBird triplicates run}
\label{tab2}
    \begin{tabular}{ccccc}    
        \toprule  
        data set & Repeats & Recall & Precision & F1-Score \\        
        \midrule            
        NIST7035\_L001 & Repeat1 & 0.9842 & 0.8926 & 0.9362\\
         & Repeat2 & 0.9842 & 0.8927 & 0.9362\\
         & Repeat3 & 0.9842 & 0.8927 & 0.9362\\
        NIST7035\_L002 & Repeat1 & 0.9845 & 0.8943 & 0.9372\\
         & Repeat2 & 0.9844 & 0.8943 & 0.9372\\
         & Repeat3 & 0.9844 & 0.8943 & 0.9372\\
        NIST7086\_L001 & Repeat1 & 0.9837 & 0.8951 & 0.9373\\
         & Repeat2 & 0.9837 & 0.8952 & 0.9374\\
         & Repeat3 & 0.9837 & 0.8951 & 0.9373\\
        NIST7086\_L002 & Repeat1 & 0.9858 & 0.8928 & 0.9370\\
         & Repeat2 & 0.9858 & 0.8928 & 0.9370\\
         & Repeat3 & 0.9858 & 0.8928 & 0.9370\\
        SRR098359 & Repeat1 & 0.9845 & 0.9012 & 0.9410\\
         & Repeat2 & 0.9845 & 0.9011 & 0.9410\\
         & Repeat3 & 0.9845 & 0.9011 & 0.9410\\
        SRR742200 & Repeat1 & 0.9870 & 0.9218 & 0.9533\\
         & Repeat2 & 0.9870 & 0.9218 & 0.9533\\
         & Repeat3 & 0.9870 & 0.9218 & 0.9533\\
        SRR098401 & Repeat1 & 0.9827 & 0.8944 & 0.9365\\
         & Repeat2 & 0.9828 & 0.8944 & 0.9365\\
         & Repeat3 & 0.9828 & 0.8943 & 0.9365\\
        ERR262997 & Repeat1 & 0.8857 & 0.8733 & 0.8794\\
         & Repeat2 & 0.8857 & 0.8732 & 0.8794\\
         & Repeat3 & 0.8857 & 0.8733 & 0.8794\\
        NA12878\_1000G & Repeat1 & 0.9624 & 0.8938 & 0.9269\\
         & Repeat2 & 0.9624 & 0.8939 & 0.9269\\
         & Repeat3 & 0.9624 & 0.8938 & 0.9269\\
        ERR091571 & Repeat1 & 0.9817 & 0.9233 & 0.9516\\
         & Repeat2 & 0.9817 & 0.9234 & 0.9516\\
         & Repeat3 & 0.9817 & 0.9233 & 0.9516\\

        \bottomrule         
        \end{tabular}
\end{table}

\subsection{Variant calling consistency}
To further assess the repeatability of the variant calling results, we conducted the triplicated variant calling runs of HummingBird pipeline on 10 samples, and the corresponding F1-scores are shown in Figure 6. results showed that HummingBird has a stable variant calling ability.

\begin{figure*}[htbp]
\centerline{\includegraphics[height=7cm,width=16cm]{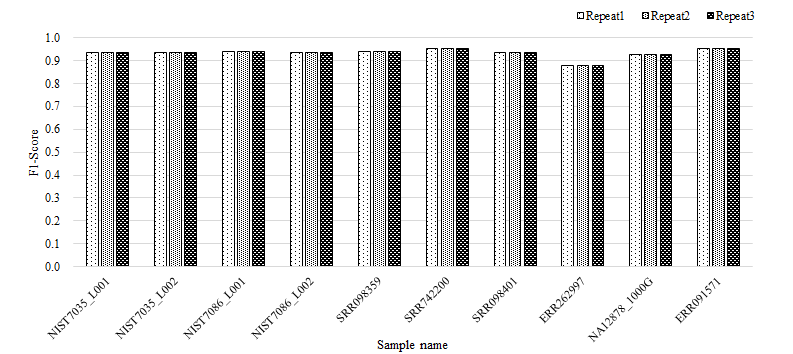}}
\caption{Consistency of variant call by HummingBird pipeline. HummingBird pipeline were run in triplicates, accuracy metrics of the triplicated variant calling runs of HummingBird pipeline were evaluated on selected samples. }
\label{fig6}
\end{figure*}

\section{Discussion and Future Directions}
In this study, we validated the efficiency of HummingBird germline variant calling pipeline by comparing it with the widely accepted BWA-GATK pipeline. High speedup and comparable accuracy support the superiority of HummingBird than BWA-GATK pipeline. Thus, it provides an alternative way of germline mutation variant calling to meet the needs of the community.

The optimization to germline mutation variant calling on standalone machine is important but not adequate. In this study, we did not test its performance in a cluster computation environment, although the pipeline can be readily implemented. With the rapid increase of human genomics data, we need to pay more effort to implement and optimize HummingBird pipeline on distributed clusters and thus bear larger scale computing.

\section*{Acknowledgements}
We thank the individuals who have made their data publicly available. We also would like to thank the public sequencing projects made the genomic data available for different researchers to explore. We thank Wei Yue, Lei Ye, Jinpeng Jiang, Yueyan Zhao, Jianhua Gao, Shiyu Han and Zhaonian Tan for their inputs during this manuscript preparation.

\end{document}